\documentclass[a4paper,12pt]{article}

\usepackage{amssymb,amsfonts,amsmath}
\usepackage{subfig,graphicx,caption}
\usepackage{hyperref}  
\usepackage{setspace}
\usepackage{fullpage}
\usepackage{authblk}    

\onehalfspace

\title{Effects of Nonlinear Coupling on Spatiotemporal Regularity}
\author[1]{Ankit Kumar}
\affil{Indian Institute of Science Education and Research (IISER) Mohali, Knowledge City, SAS Nagar, Sector 81, Manauli PO 140306, Punjab, India}

\begin{document}
\date{}
\maketitle

\begin{abstract}
  In this work we investigate the spatiotemporal behaviour of lattices
  of coupled chaotic logistic maps, where the coupling between sites
  has a nonlinear form. We show that the stable range of the
  spatiotemporal fixed point state is significantly enhanced for
  increasingly nonlinear coupling. We demonstrate this through
  numerical simulations and linear stability analysis of the
  synchronized fixed point. Lastly, we show that these results also
  hold in coupled map lattices where the nodal dynamics is given by
  the Gauss Map, Sine Circle Map and the Tent Map.

\end{abstract}


\section{Introduction}
  Coupled Map Lattices (CML) is a class of models that have been used
  to describe many spatially extended complex systems
  \cite{book01,book03}, ranging from ecological networks
  \cite{ref07,ref08,ref09} to nematic liquid crystals \cite{ref05}.
  The basic ingredients of a CML model is the local dynamics at the
  sites of the lattice and the coupling interaction between subsets of
  the sites. In recent years researchers have extensively explored the
  effect of different coupling topologies on the dynamics of coupled
  dynamical systems and found remarkable phenomenon such as
  spatiotemporal synchronization. Network topology has been considered
  the key to obtaining a stable synchronization manifold in coupled
  chaotic systems \cite{ref03,ref02}. However the effect of the form of the coupling
  function on spatiotemporal patterns has not gained much research
  attention. In this work we address this issue and show that the
  nonlinear coupling forms have a pronounced effect on spatio-temporal
  synchronization in coupled map systems. Our central results is that
  nonlinear coupling stabilizes the synchronization manifold, even for
  regular networks where average path length is too high to sustain a
  stable synchronization manifold.

 \textit{Model: } We consider $N$ chaotic maps on an ring. The sites are denoted by the
 index $i=1,2\dots N$, where $N$ is the system size. Each node is
 coupled to its two nearest neighbors i.e. the $i^{th}$ site is
 connected to site $(i+1)^{th}$ and $(i-1)^{th}$ node.  The general time
 evolution equation of $i^{th}$ node is given by
 \begin{equation}
 x^{i}_{n+1} = (1-\epsilon)f(x^{i}_n) + \frac{\epsilon}{2K}
 \sum_{j=1}^{K} (g(x^{i-j}_n) + g(x^{i+j}_n) )
 \label{evolution}
 \end{equation}

 In this work we will consider $K=1$, namely each site couples to two
 nearest neighbours. In the language of networks $i$ represents a node
 of degree $2$ in the network. The function $f(x)$ gives the local
 dynamics. To begin with, this is chosen to be the prototypical chaotic
 map, the logistic map: $ x_{n+1} = rx_n(1-x_n)$, where the
 nonlinearity parameter $r$ is chosen to be $4$. The strength of
 coupling is denoted by $\epsilon$, and the function $g(x)$ is the
 coupling function. The $(1-\epsilon)$ term weighting the local map
 $f(x)$ keeps the system bounded by confining the dynamics of the nodes
 in the interval $[0:1]$.

 The central focus of this work is to explore the behaviour of this
 system under coupling of the form:
 \begin{equation}
 g(x)= x^q
 \label{power}
 \end{equation}
 where $q=1,2,3, \dots \infty$. Since $x\leq 1$, in this class of examples, increasing power of $x$ yield smaller
 magnitude of  the coupling term. 

 In the sections below we will present our
 results from extensive numerical simulations and stability analysis,
 describing the effects of varying $q$ in the coupling function, on
 spatiotemporal patterns. Notably, we will show that increasingly
 nonlinear coupling forms yield larger stable ranges for the
 spatiotemporal fixed point.

\section*{Numerical Results}

\begin{figure}[]
\subfloat[Linear coupling]{\label{subfig: bif1} \includegraphics[width=0.5\textwidth, height=70mm]{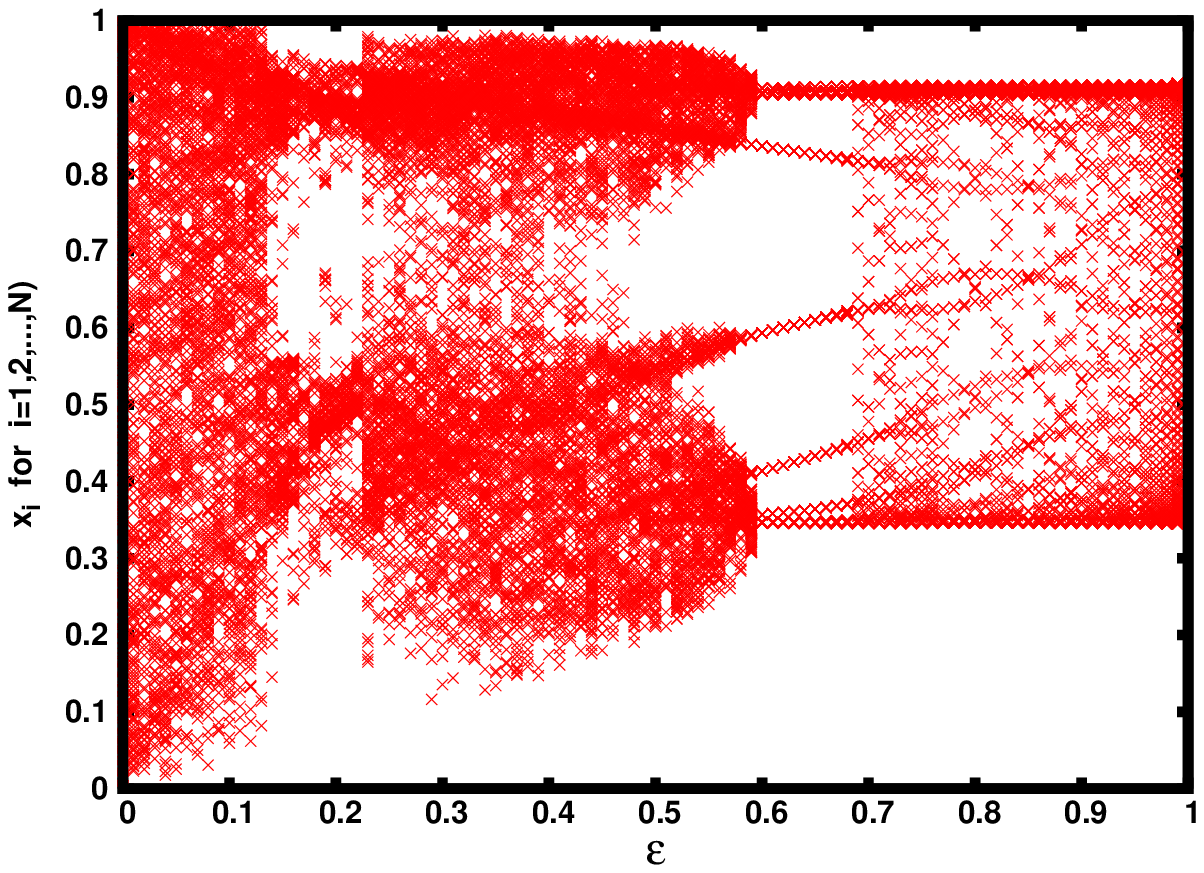}}
\subfloat[Quadratic coupling]{\label{subfig: bif2} \includegraphics[width=0.5\textwidth, height=70mm]{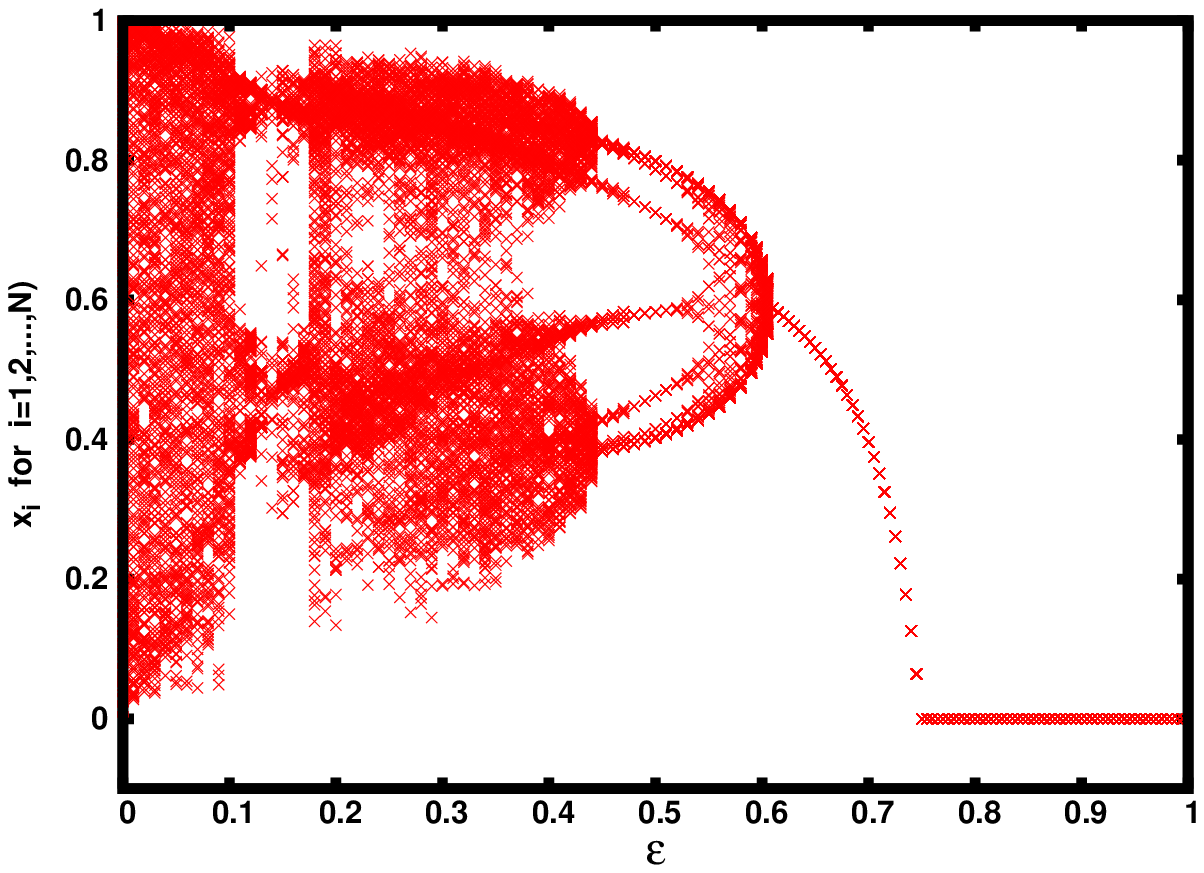}}\\
\subfloat[Linear coupling]{\label{subfig: bif1_1} \includegraphics[width=0.5\textwidth, height=80mm]{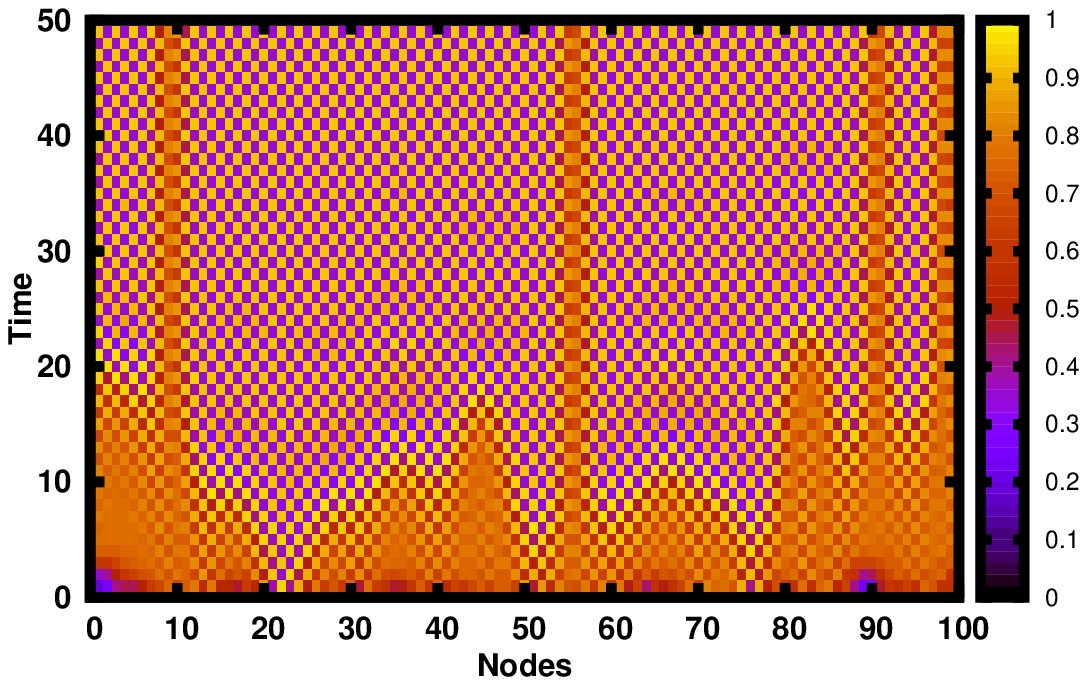}}
\subfloat[Quadratic coupling]{\label{subfig: bif2_1} \includegraphics[width=0.5\textwidth, height=80mm]{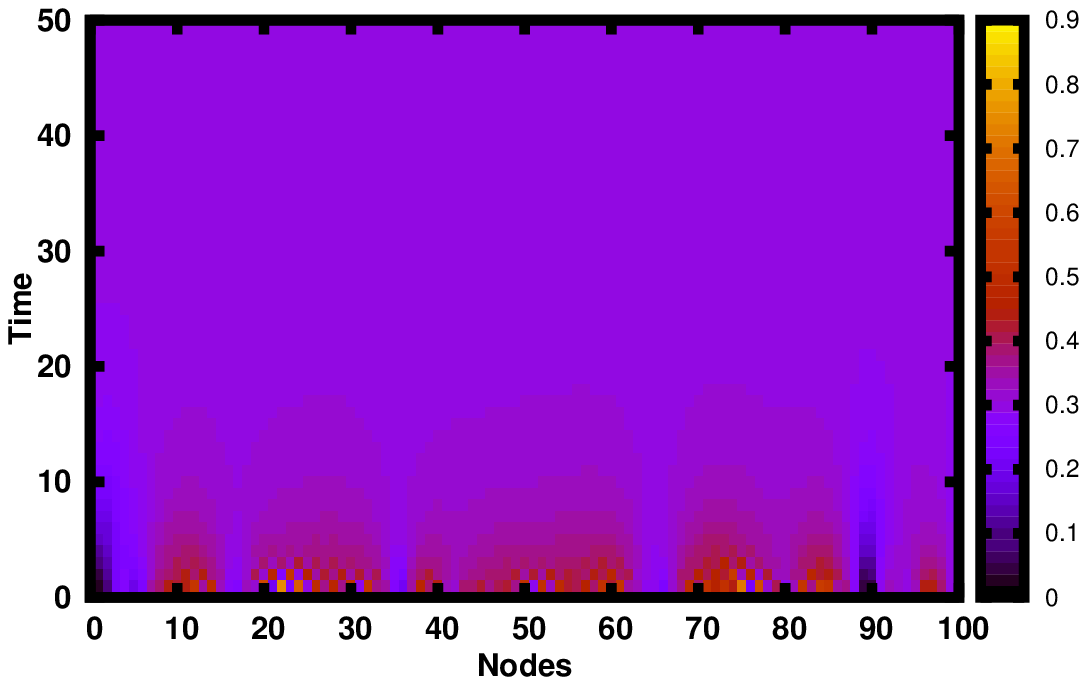}}\\
\caption{ Subfigure \textbf{(a) and (b):} Bifurcation diagram showing values of $x_{n}^i$ with respect
  to coupling strength $\epsilon$, for coupled logistic maps with
  linear coupling ($q=1$ in Eqn.~1-2) and quadratic coupling ($q=2$ in
  Eqn.~1-2). Here the size of the lattice is $N=100$. In the figure we
  plot $x_n^i(i=1,...,100)$ over $n=1,...,10$ iterations (after a
  transience time of $5000$) for ten different initial conditions.\\
  Subfigure \textbf{(c) and (d):} Showing spatio-temporal dynamics of system for coupling strength $\epsilon=0.72$ 
 for coupled logistic maps with linear coupling and quadratic coupling.}
 \label{fig: bif5}
\end{figure}

First we will demonstrate numerically that nonlinear coupling forms
enhance spatiotemporal regularity in coupled maps. The numerical
results here have been obtained for large set of random initial
conditions $(500-1000)$, with network size $N$ ranging from $10$ to
$1000$.  Figure-\ref{subfig: bif1} displays the bifurcation diagram for the linear coupling with
strictly nearest neighbors case.

 It is evident here that linear coupling scheme with strictly regular
 connections does not show any stable spatiotemporal manifold in
 system. To quantify the synchronization we define the range of synchronization as $R=(1-
 \epsilon_{sync})$. Here $\epsilon_{sync}$ is the value of $\epsilon$ where
 synchronization manifold starts getting stabilized. In case of linear coupling $R$ is zero.

 We start with the effect of the quadratic coupling form ($g(x)= x^2$)
 on the dynamics of coupled maps.  Figure-\ref{subfig: bif2} displays the
 bifurcation diagram for quadratic coupling, clearly showing the
 enhancement of the spatiotemporal fixed point regime in system.
 Notice that the invariant synchronization manifold is an $\epsilon$
 dependent curve.

\begin{figure}
 \subfloat[$x^3$ coupling]{\includegraphics[width=0.5\textwidth, height=60mm]{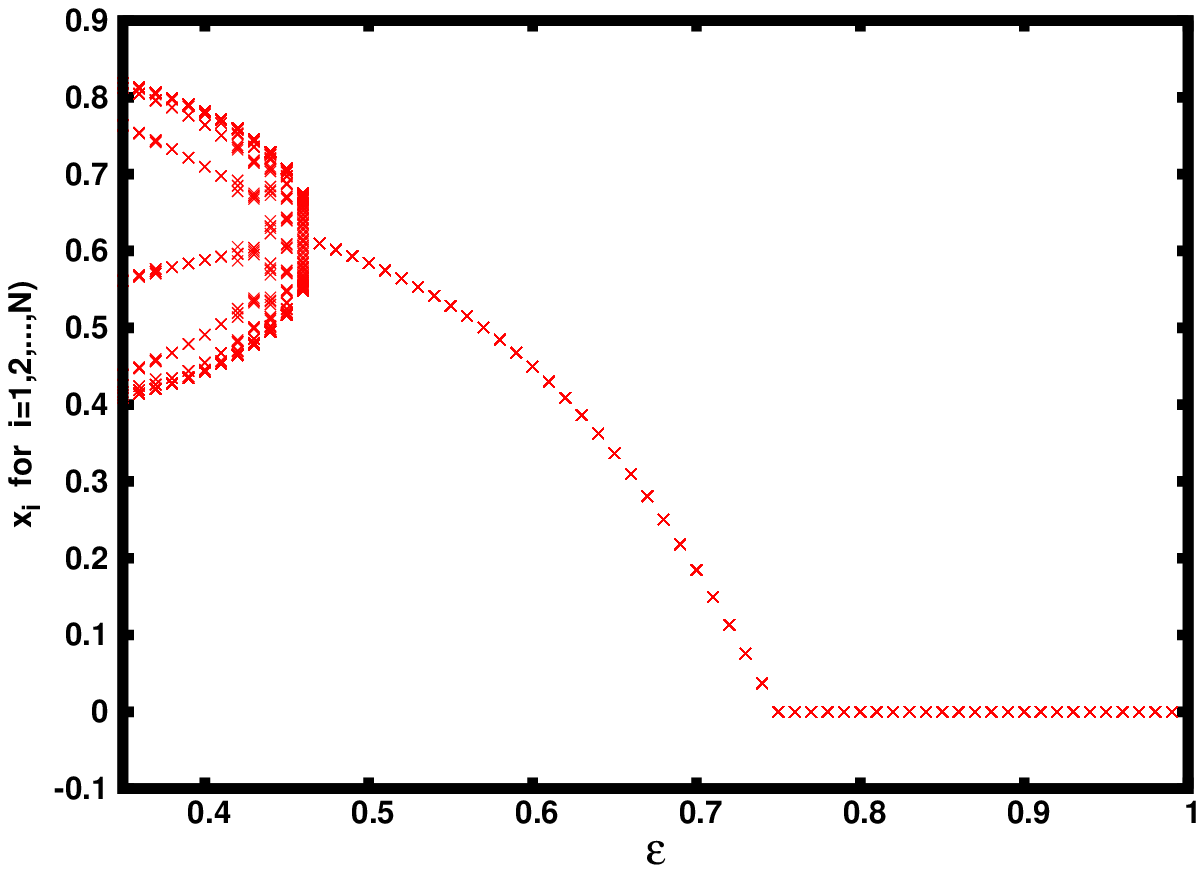}}
 \subfloat[$x^4$ coupling]{\includegraphics[width=0.5\textwidth, height=60mm]{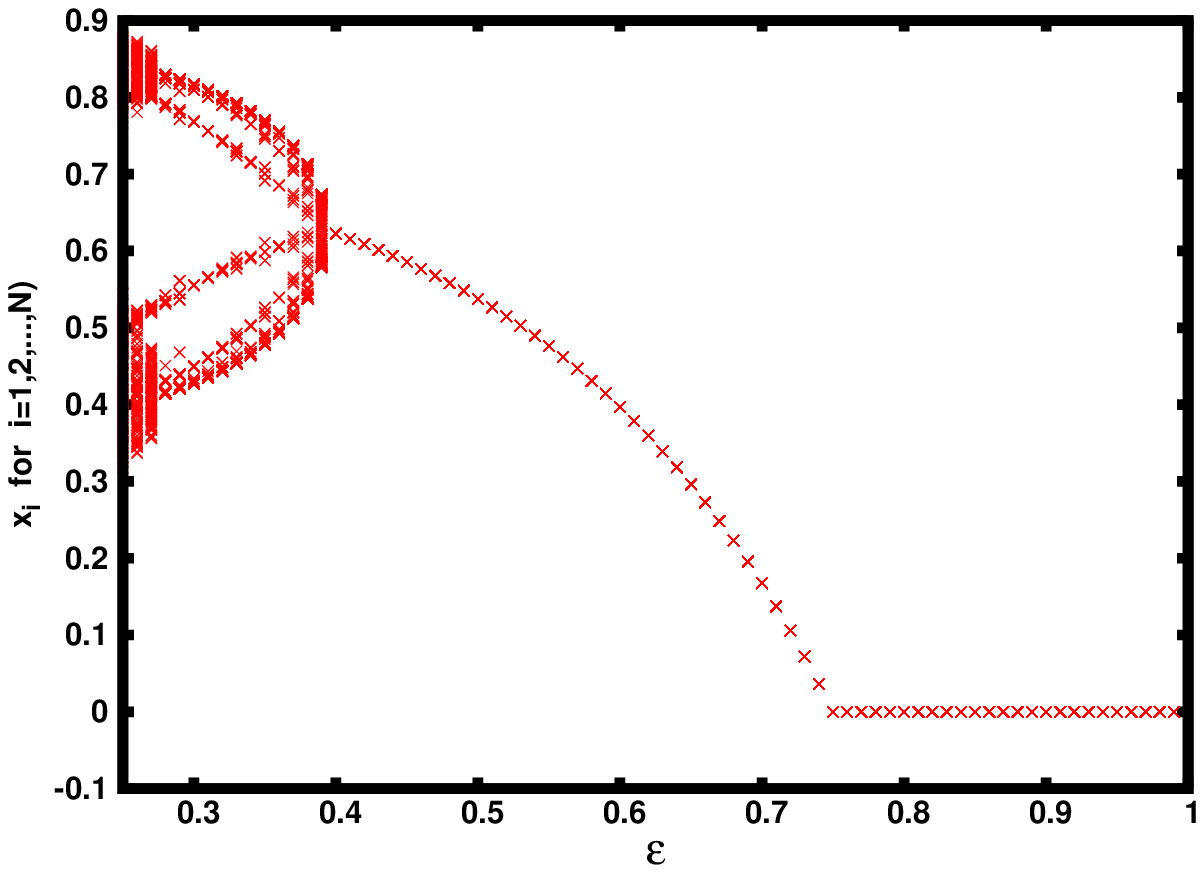}}\\
 \caption{Bifurcation diagram showing values of $x_{n}^i$ with respect
  to coupling strength $\epsilon$, for coupled logistic maps with coupling form $g(x)= x^3$ and $g(x)=x^4$.
  This figure shows how range of spatiotemporal fixed points increases as we increase the nonlinearity in coupling form.
  Here the size of the lattice is $N=100$. In the figure we
  plot $x_n^i(i=1,...,100)$ over $n=1,...,10$ iterations (after a
  transience time of $5000$) for ten different initial conditions.}
 \label{fig: bif6}
\end{figure}

\begin{figure}
\centering
\includegraphics[width=0.8\textwidth, height=80mm]{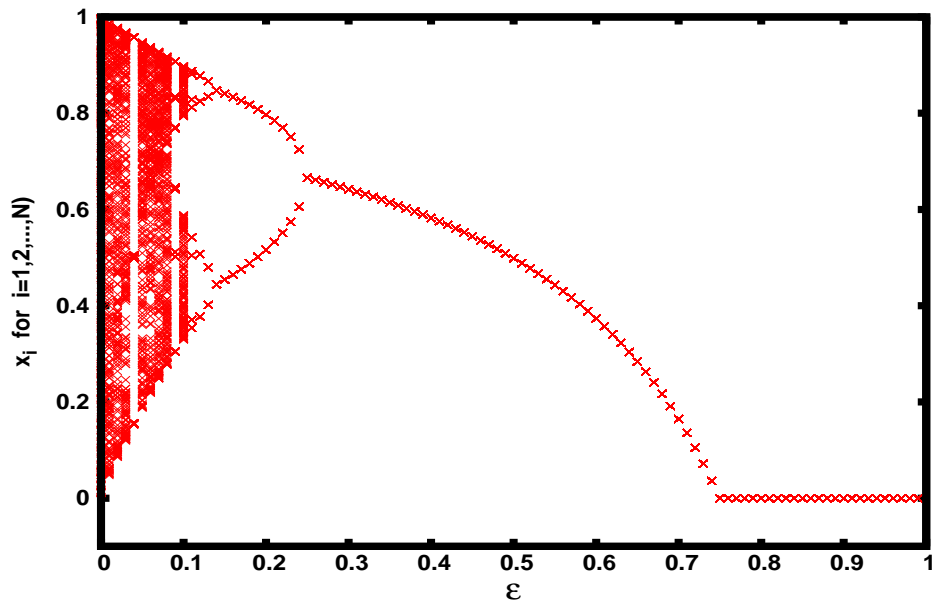}
\caption{Inverted Logistic map: Bifurcation diagram showing values of $x_{n}^i$ with respect
  to coupling strength $\epsilon$, for coupled logistic maps with
  $x^{100}$ coupling ($q=100$ in Eqn.~1-2).}
\label{fig: bif71}
\end{figure}

 We then go on to study the system with coupling function $g(x)=x^q$
 with increasing $q$. Our central observation is that the range of the
 spatiotemporal fixed point, $R$, is dependent on $q$. As we increase
 the nonlinearity in coupling we get a larger range of
 synchronization (figure-\ref{fig: bif6}, \ref{fig: bif5}). Also, The stable range of the spatiotemporal fixed point
 soon converges to a maximum $R_{max}$ as we increases $q$ further. It is worth noting that
 when we take large value of $q$, the individual nodes follow the dynamics of synchronized logistic
 map, with an inverted bifurcation sequence arising from a rescaled nonlinearity parameter (figure:\ref{fig: bif71}).

\begin{figure}[]
\subfloat[Linear coupling]{\label{subfig: bif1_1} \includegraphics[width=0.5\textwidth, height=80mm]{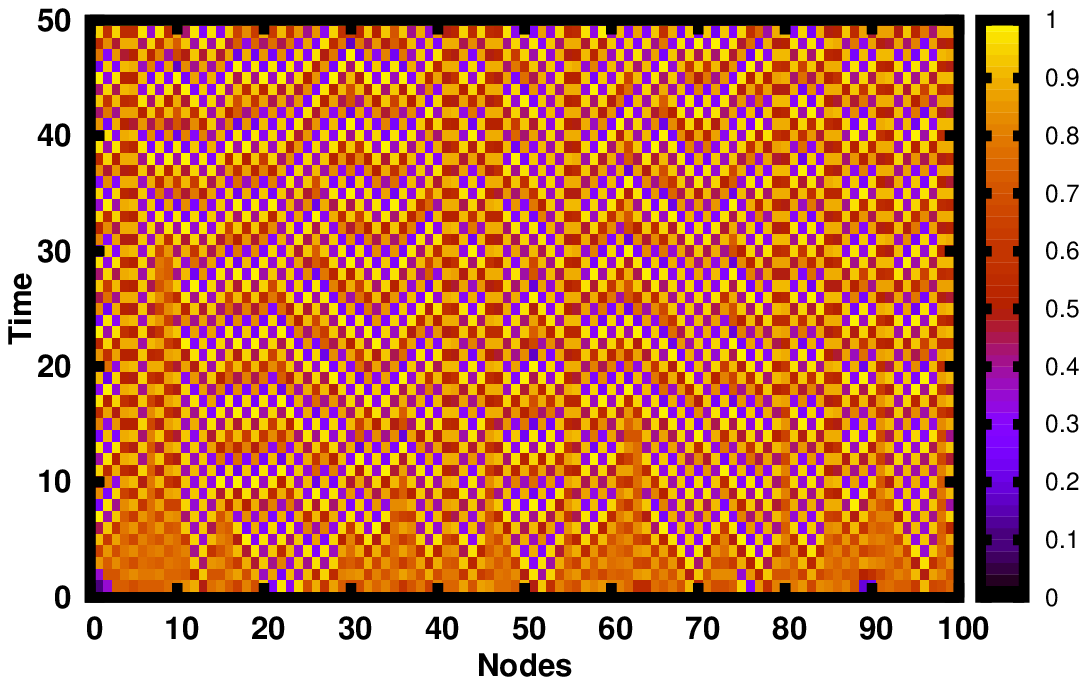}}
\subfloat[Quartic coupling]{\label{subfig: bif2_1} \includegraphics[width=0.5\textwidth, height=80mm]{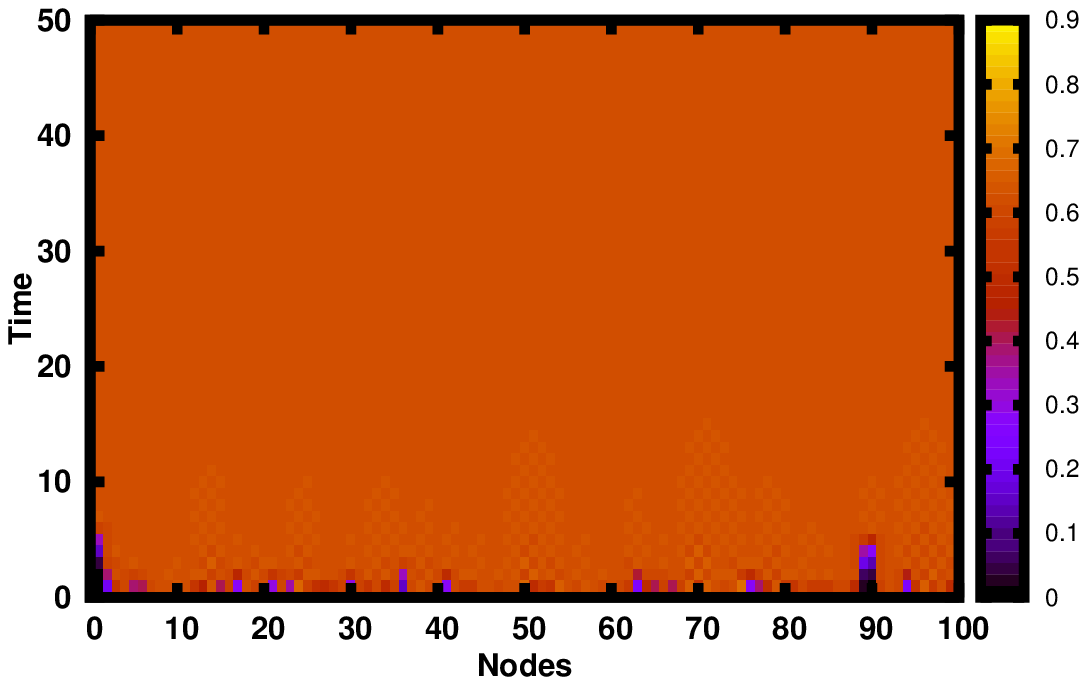}}\\
\caption{ Spatio-temporal dynamics of system for coupling strength $\epsilon=0.42$ 
 for coupled logistic maps with linear coupling and Quartic ($x^4$) coupling.}
 \label{fig: bif5}
\end{figure}

 We find the range of the spatiotemporal fixed point, numerically, for
 different $q$. The results are displayed in Figure-\ref{fig:
  bif7}. Further, we obtain the functional dependence of the coupling
 strength at which the spatiotemporal fixed point gains stability,
 denoted by $\epsilon_{sync}$, on $q$.  It is clear that
 $\epsilon_{sync}$ converges to a minimum value with increasing $q$,
 namely $\lim \limits_{q \to \infty}\epsilon_{sync} = \epsilon_{min}$.

\begin{figure}
\subfloat[Synchronization Range]{\includegraphics[width=0.5\textwidth, height=75mm]{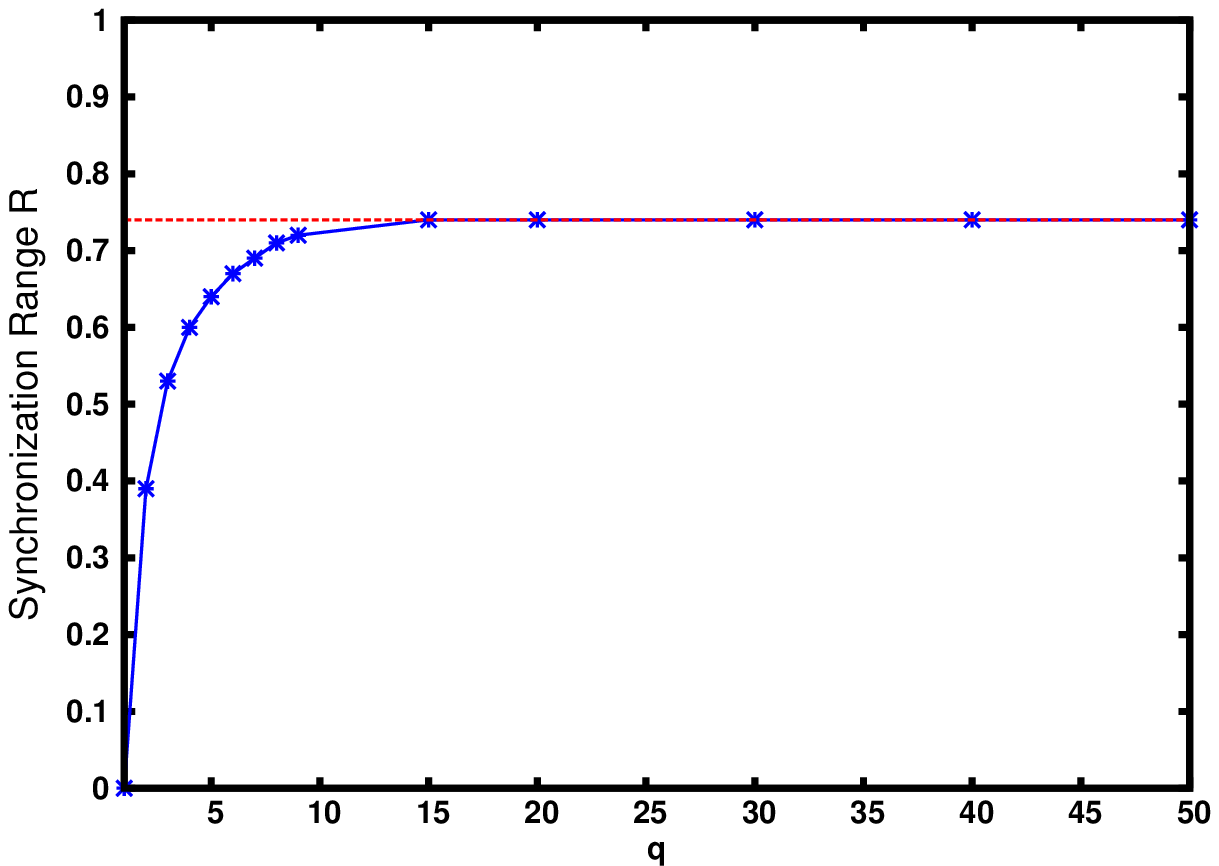}}
\subfloat[Least square fitting ]{\includegraphics[width=0.5\textwidth, height=75mm]{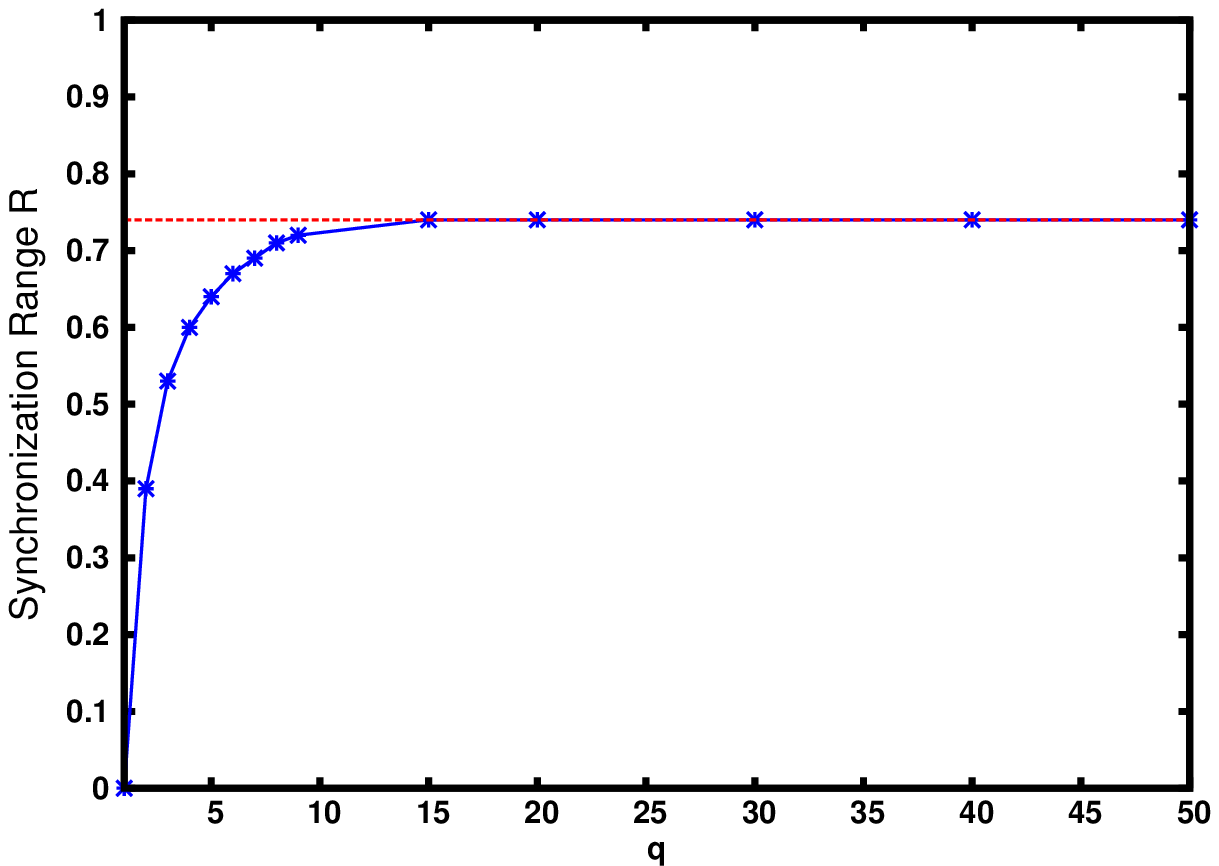}}
\caption{Characterizing the dependence of Range on $q$. (a) Numerically calculated ranges R for many q's. 
(b) Best fitting for $\epsilon_{sync}$ vs $q$ }
\label{fig: bif7}
\end{figure}

In order to understand the coupling strength necessary to stabilize
the spatiotemporal fixed point, we attempted to fit the values of
$\epsilon_{sync}$ obtained for increasing values of $q$, to different
functional forms. We found reasonably good fit to the form:
\begin{equation}
\epsilon_{sync} = \frac{a}{q^b} +c
\end{equation}
with parameters values $a=0.977 \pm 0.063$, $b=1.1415 \pm 0.078$ and
$c=0.251 \pm 0.004$.

In the next section we will analyse the system to gain more
understanding of the effect of nonlinear coupling forms on
spatiotemporal regularity.

\section*{Analysis}

Mathematically, the synchronization manifold is the state of system when all nodes follow same trajectory i.e. 
$x^1=x^2=\dots=x^N$, and for the temporal fixed point we must have: $x^{i}_{n+1} = x^{i}_{n} = x^*$ for
$i=1,\dots,N$, where $x^*$ is the spatiotemporal fixed point. Combinedly, these two equations
result following one equation which can be solved to compute spatiotemporal fixed point.
\begin{align}
x^* &= (1-\epsilon ) r x^* (1-x^*) + \frac{\epsilon}{2}\sum_{j=1}^{2}(x^*)^{q} \\
 \epsilon x^q &- (1-\epsilon)rx^2 +(r-r\epsilon -1)x = 0  \label{eqn: eq1}  
\end{align}
For linear coupling, i.e. $q=1$, and for $r=4$, solution of above equation-\ref{eqn: eq1} are
\begin{align*}
x = 0 \quad or \quad x &= 0.75 \\
\end{align*}
So linear coupling gives two $\epsilon$ independent fixed points.

\begin{figure}
\centering
\includegraphics[width=0.8\textwidth, height=80mm]{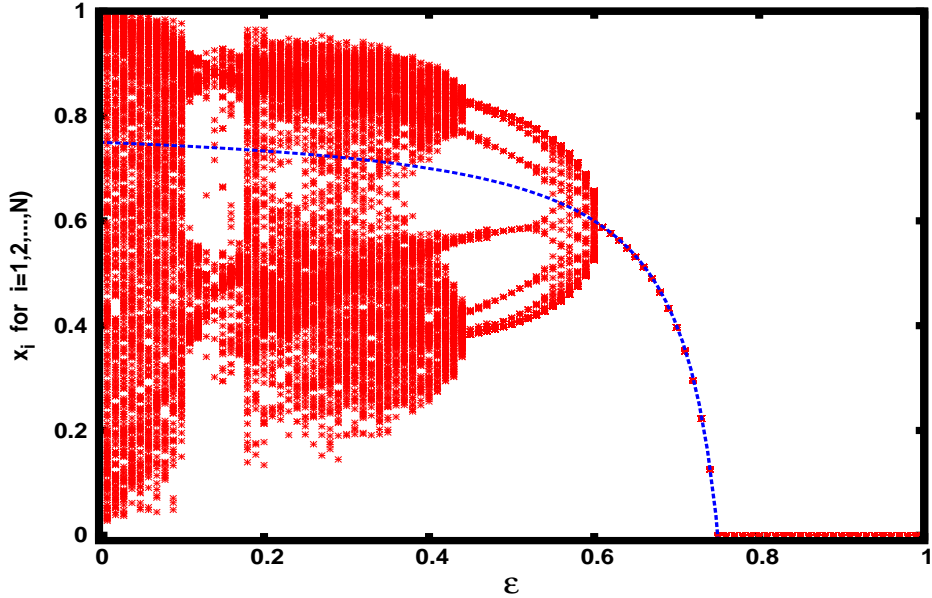}
\caption{Spatiotemporal fixed points for the case of quadratic
  coupling. Here the points marked with crosses are obtained
  numerically, and the dashed line is the analytical solution.}
\label{fig: bif8}
\end{figure}

In case of quadratic coupling ($q=2$), equation-\ref{eqn: eq1} provides following two  solutions:
\begin{align*}
x = 0 \quad or \quad x &= \frac{-3 +4\epsilon }{-4 + 5\epsilon } \\
\end{align*}
In this case, one of the solutions is coupling strength, $\epsilon$,
dependent. We compare this solution with numerical simulations
(figure-\ref{subfig: bif2}) and get exact match as it is clearly shown in figure-\ref{fig: bif8}.

 By following the same approach as above, we can analytically find the
 spatiotemporal fixed points for any $q$. 
 
 Note that the limit of $q$ tending to infinity is interesting because we have $0 \leq x_n \leq 1$ for all $n$,
 that implies $\lim \limits_{q \to \infty}(x_{n})^q \backsimeq 0$. So evolution equations of system can be written as
 $$x^{i}_{n+1} = (1-\epsilon )rx^i_n(1-x^i_n) + 0$$
 This equation is nothing but the \textit{inverted logistic map} with effective nonlinearity parameter $r(1-\epsilon)$.
 Bifurcation diagram is shown in figure:\ref{fig: bif71}.

\subsubsection*{Linear stability analysis of the spatiotemporal fixed points}

In the previous section we calculated the fixed points for the
nonlinear coupling case.  Now we will check the stability of these
fixed points using standard \textit{linear stability analysis}
\cite{ref04} .

\begin{figure}[h]
\centering
\subfloat[Fixed Point at Zero]{\includegraphics[width=0.5\textwidth, height=60mm]{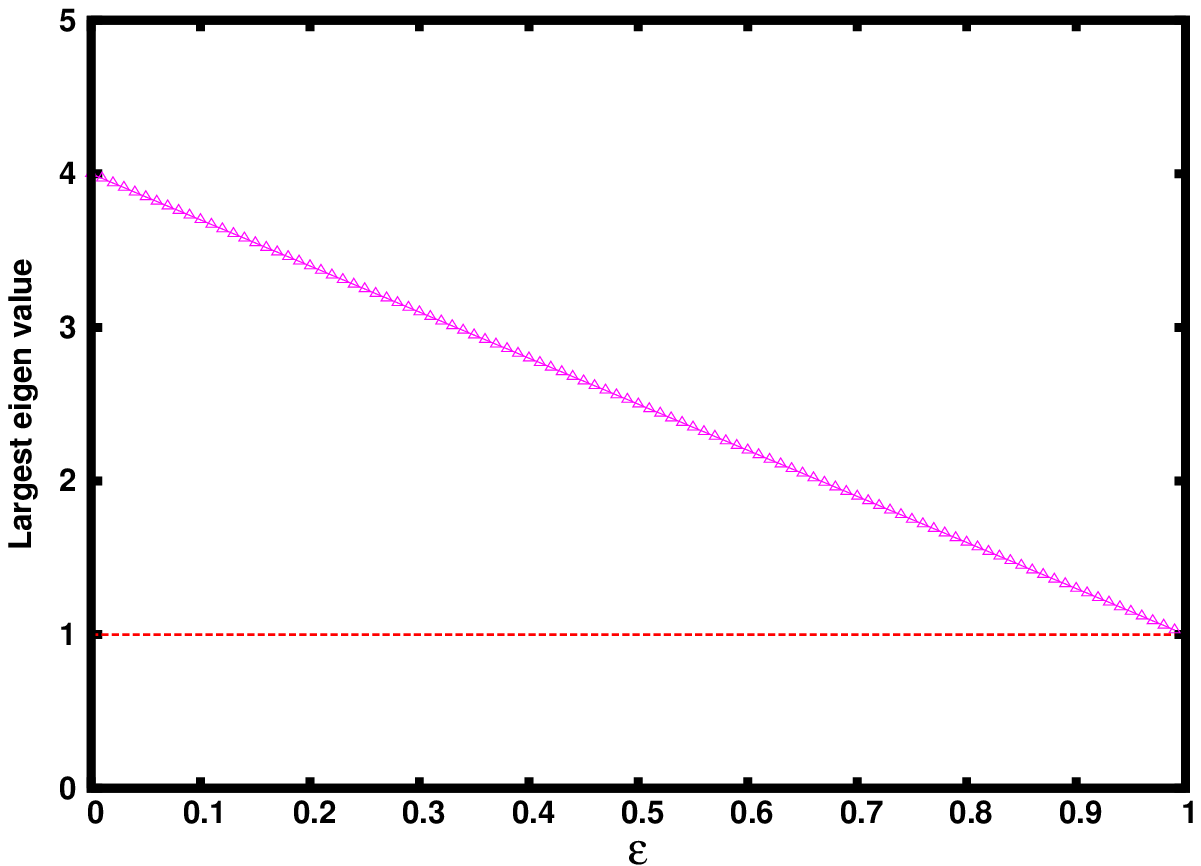}}
\subfloat[Nonzero Fixed Point]{\includegraphics[width=0.5\textwidth, height=60mm]{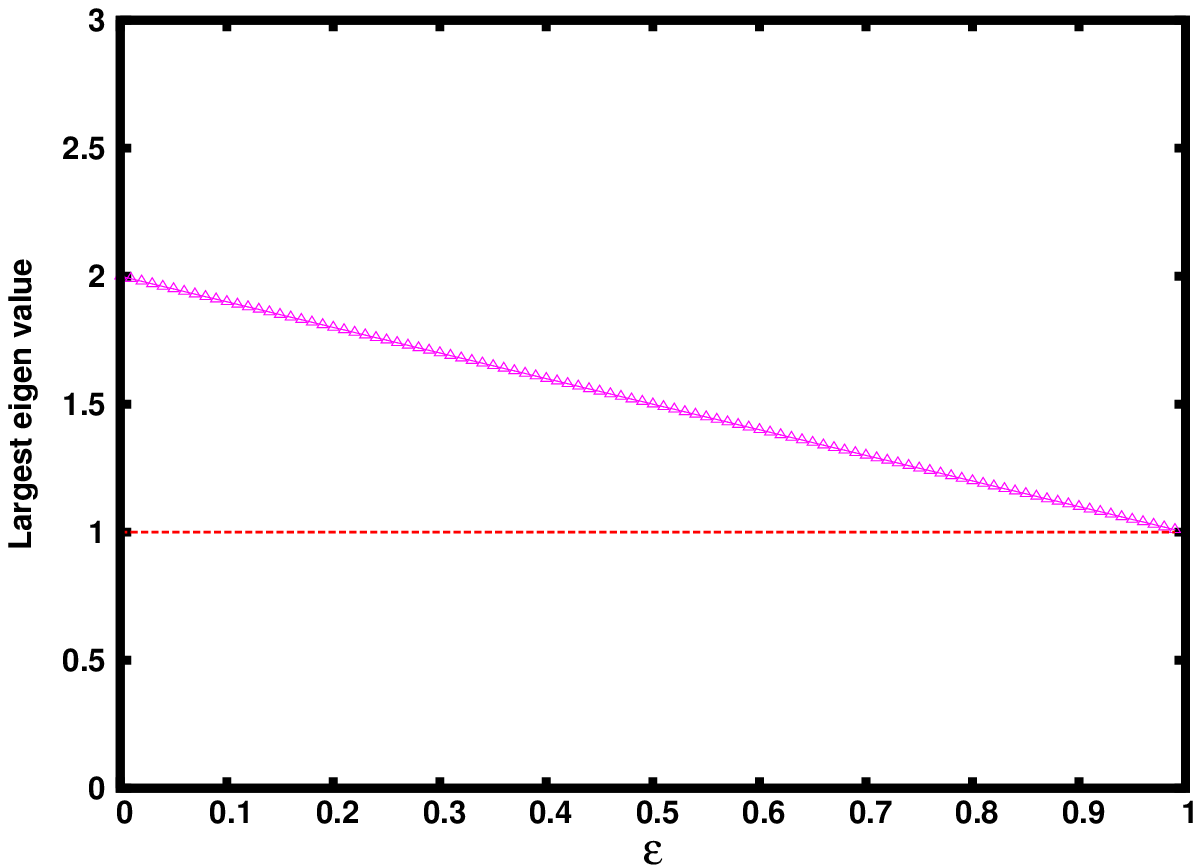}}
\caption{ Magnitude of largest eigenvalue vs $\epsilon$ plot for linear coupling.}
\label{fig: eig1} 
\end{figure}

\textbf{Linear coupling} : As calculated above, there exists two
spatiotemporal fixed points in the linear coupling case, and
simulations showed that both are unstable. So we expect that the maximum magnitude of eigenvalues of Jacobian matrix, calculated around
these two fixed points, should be {\em greater than $1$}. In order to
check this, we analytically find the eigenvalue spectrum of the
Jacobian matrix \cite{book02} and plot the magnitude of largest eigenvalue, with respect to coupling strength $\epsilon$. The results
are displayed in Figure \ref{fig: eig1}.

Figure-\ref{fig: eig1} exactly matches the numerical results, and shows that
there is no stable spatio-temporal fixed point state as the magnitude of largest eigenvalue is always greater than $1$ for both fixed points.

\begin{figure}[h]
 \centering
 \subfloat[Fixed Point at Zero]{\includegraphics[width=0.5\textwidth, height=60mm]{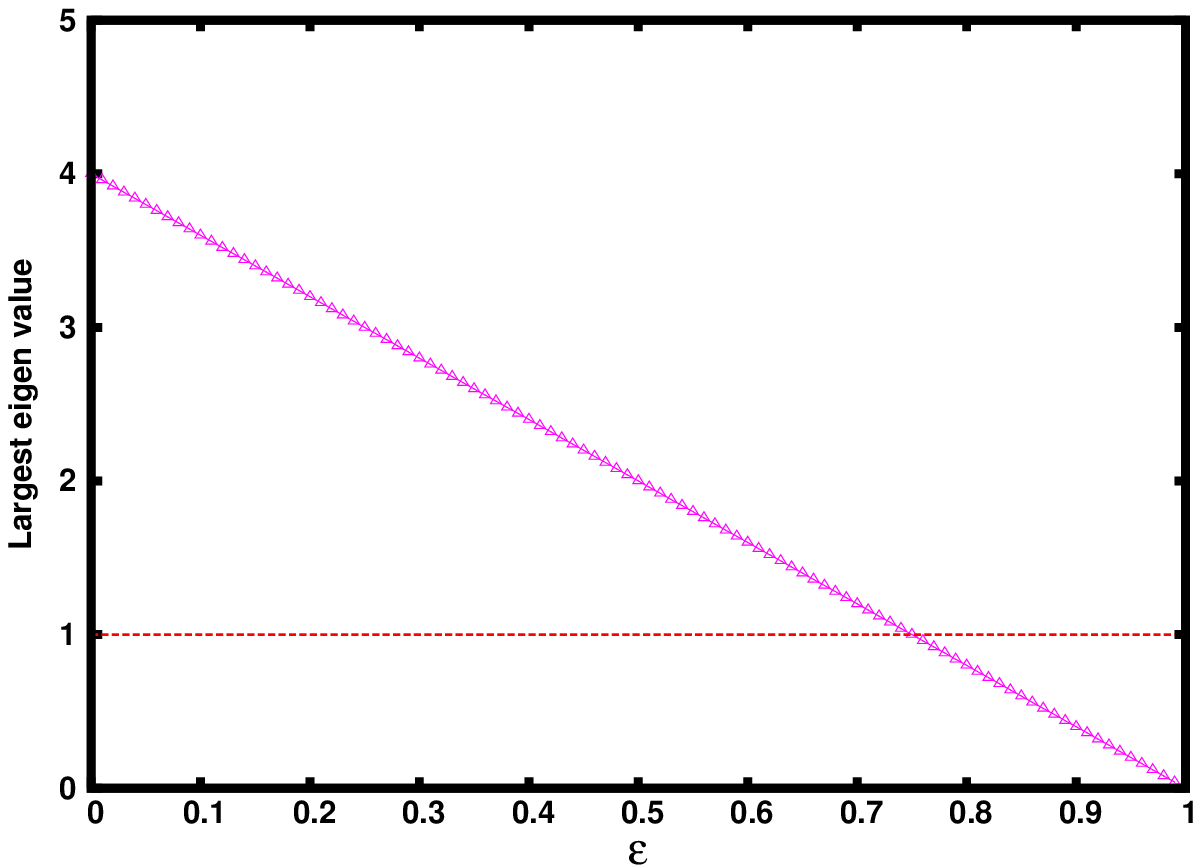}}
 \subfloat[Nonzero Fixed Point]{\includegraphics[width=0.5\textwidth, height=60mm]{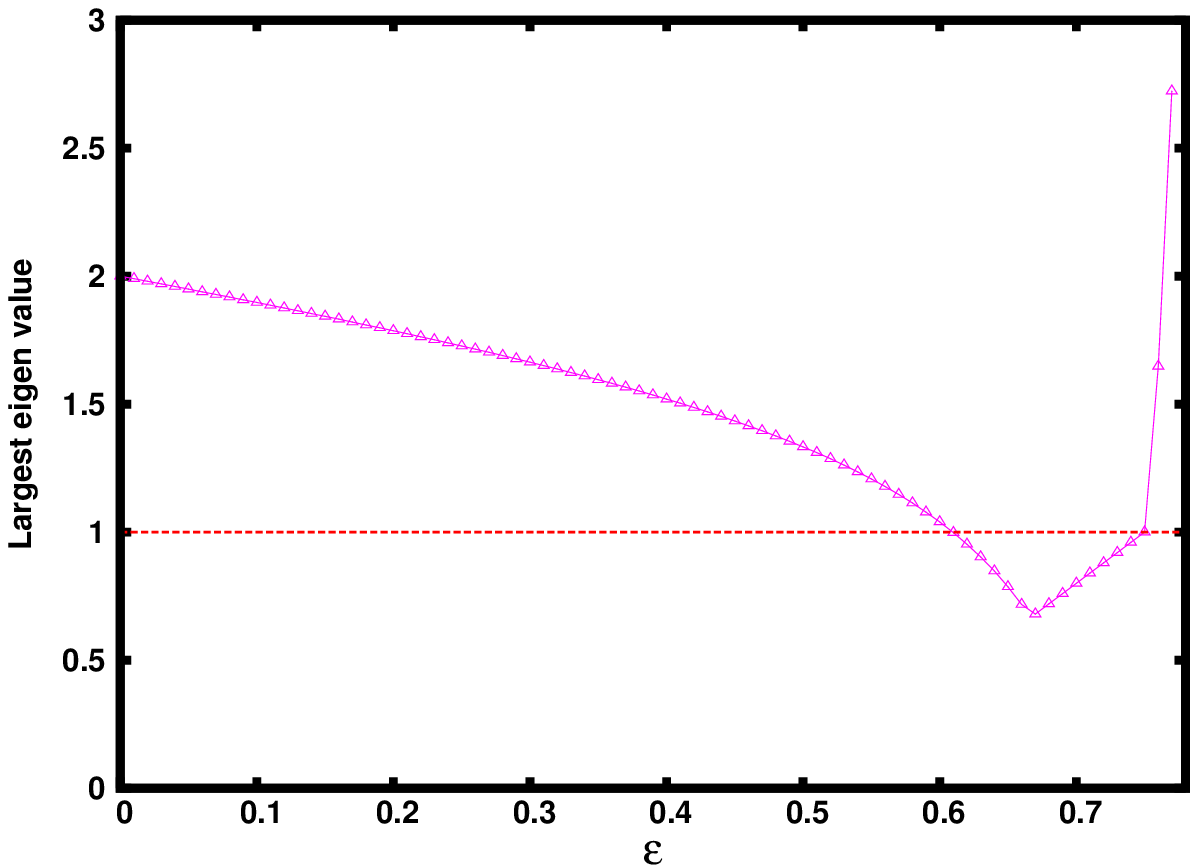}}
 \caption{ Magnitude of largest eigenvalue vs $\epsilon$ plot for quadratic coupling.}
 \label{fig: eig2} 
\end{figure}

 For the quadratic coupling case ($q=2$) results are displayed in the
 Figure \ref{fig: eig2}.  It is evident here that the fixed point at
 zero is stable for $0.75 \leq \epsilon \leq 1$ because the maximum magnitude of eigenvalues
 is smaller than $1$ for this $\epsilon$ range.  On the
 other hand the non-zero fixed point is stable for $0.61 \leq \epsilon
 \leq 0.75$.  These ranges are exactly as observed in the numerical
 simulations (cf.  bifurcation diagram in Figure \ref{subfig: bif2}).

 So it is clear from the analysis above that we can analytically
 predict the stability of the spatiotemporal fixed points,
 qualitatively as well as quantitatively.
\section*{Generality of the Results}

In order to examine the range of applicability of the phenomena
observed above, we will implement the same coupling scheme in coupled
map lattices with different local dynamics. In particular we will
demonstrate below how nonlinear coupling yields enhanced stable
regimes for the spatiotemporal fixed point state, in a system whose
local evolution is given by the Gauss Map, Sine-Circle map and by the
Tent Map.

The {\em Gauss Map} is given as follows:

$$ x_{n+1} = e^{-\alpha x_{n}^2} +\beta$$
\begin{figure}[h]
\centering
\subfloat[Linear coupling]{\includegraphics[width=0.5\textwidth, height=55mm]{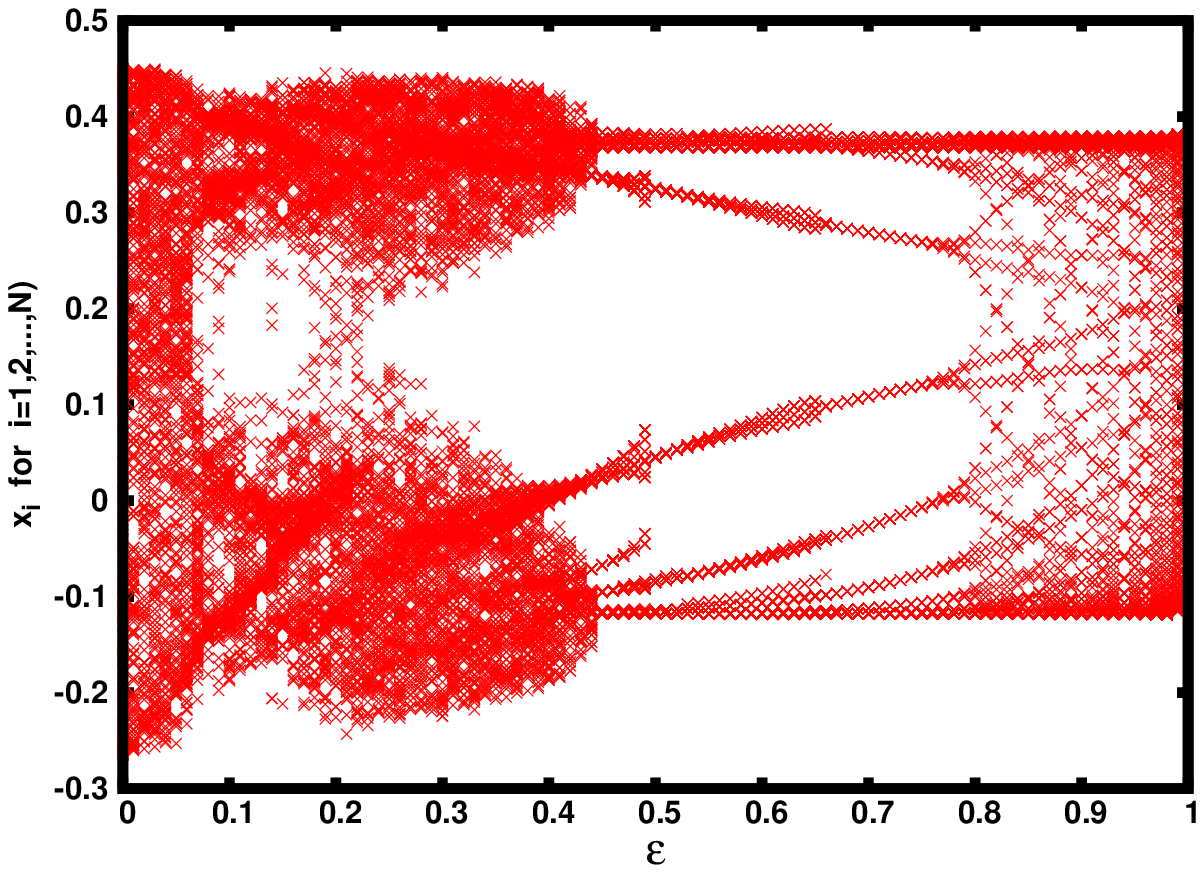}}
\subfloat[Quadratic Coupling]{\includegraphics[width=0.5\textwidth, height=55mm]{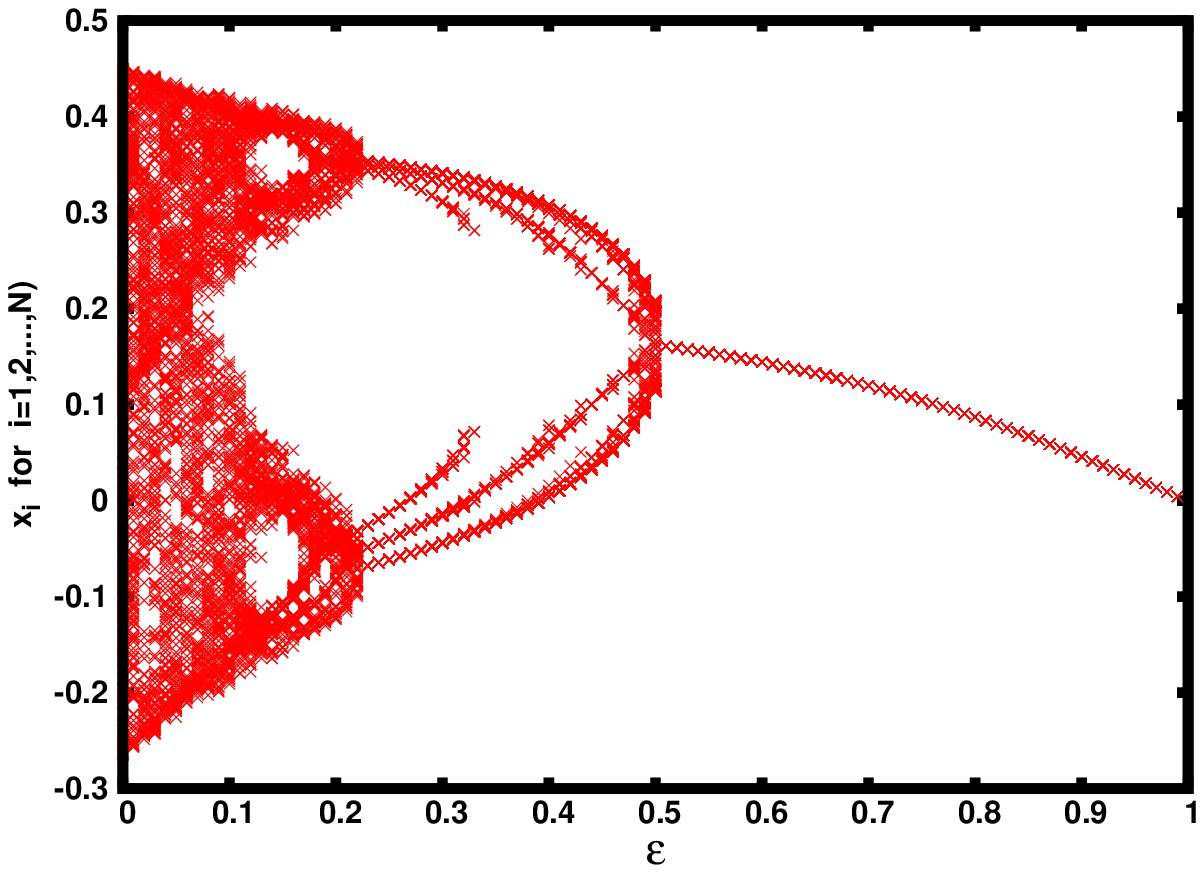}}
\caption{Bifurcation diagram showing values of $x_{n}^i$ with respect
  to coupling strength $\epsilon$, for coupled Gauss maps with
  linear coupling ($q=1$ in Eqn.~1-2) and quadratic coupling ($q=2$ in
  Eqn.~1-2). Here the size of the lattice is $N=100$. In the figure we
  plot $x_n^i(i=1,...,100)$ over $n=1,...,10$ iterations (after a
  transience time of $5000$) for ten different initial conditions.}
\label{fig: bif13}
\end{figure}

It shows chaotic dynamics for parameter values
$(\alpha,\beta)=(-6.2,0.55)$. Now we take this chaotic map as our
local dynamics and simulate the dynamics of system for quadratic
($g(x) =x^2$) as well as linear coupling. The results are presented in
Figure \ref{fig: bif13}. It is clear from the figure that we get a
stable range for the spatiotemporal fixed point state when the
coupling is nonlinear, while linear coupling does not give any stable
range.

The {\em Tent Map} is given as follows:
$$x_{n+1}=  \begin{cases} 
      rx,       & x\leq \frac{1}{2} \\
      r(1-x),   & \frac{1}{2} \leq x\leq 1 \\
   \end{cases}
$$
\begin{figure}[h]
\centering
\subfloat[Linear coupling]{\label{}\includegraphics[width=0.5\textwidth, height=50mm]{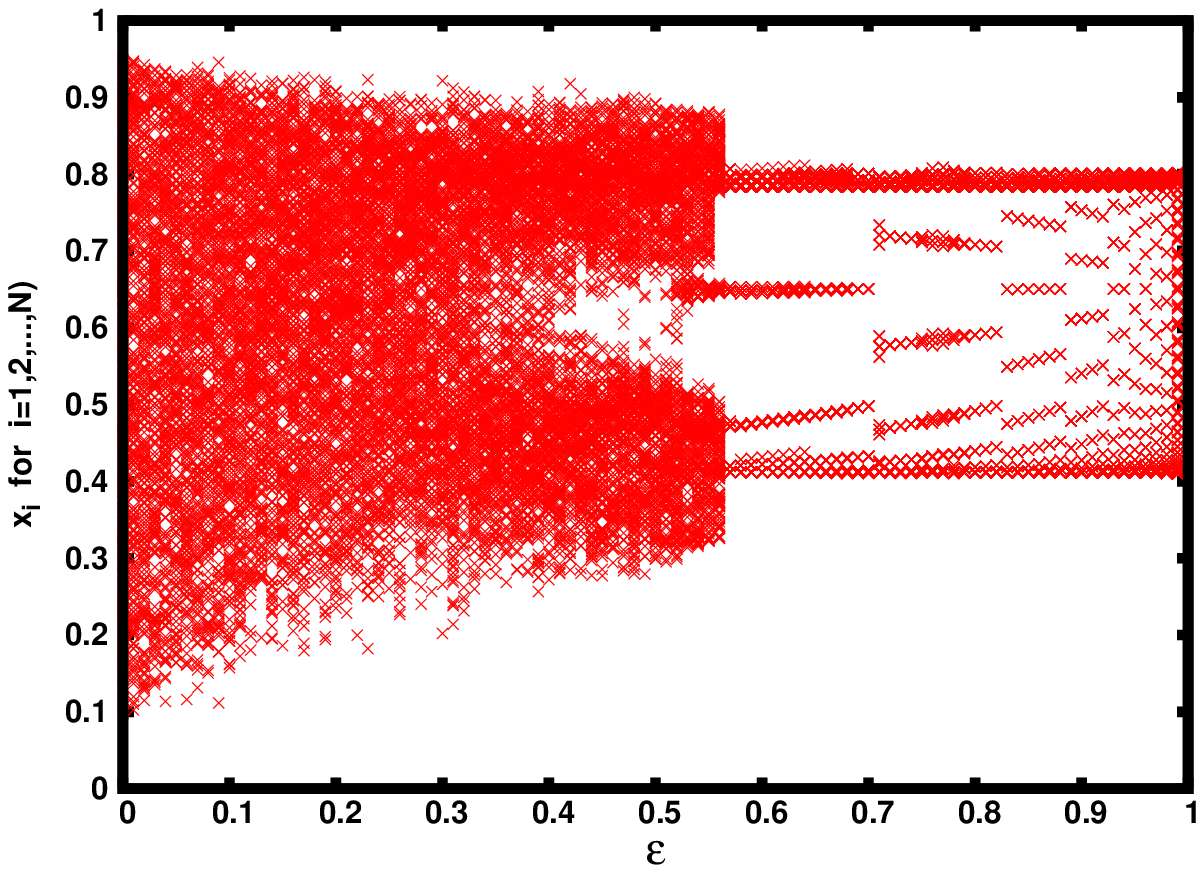}}
\subfloat[Quadratic Coupling]{\includegraphics[width=0.5\textwidth, height=50mm]{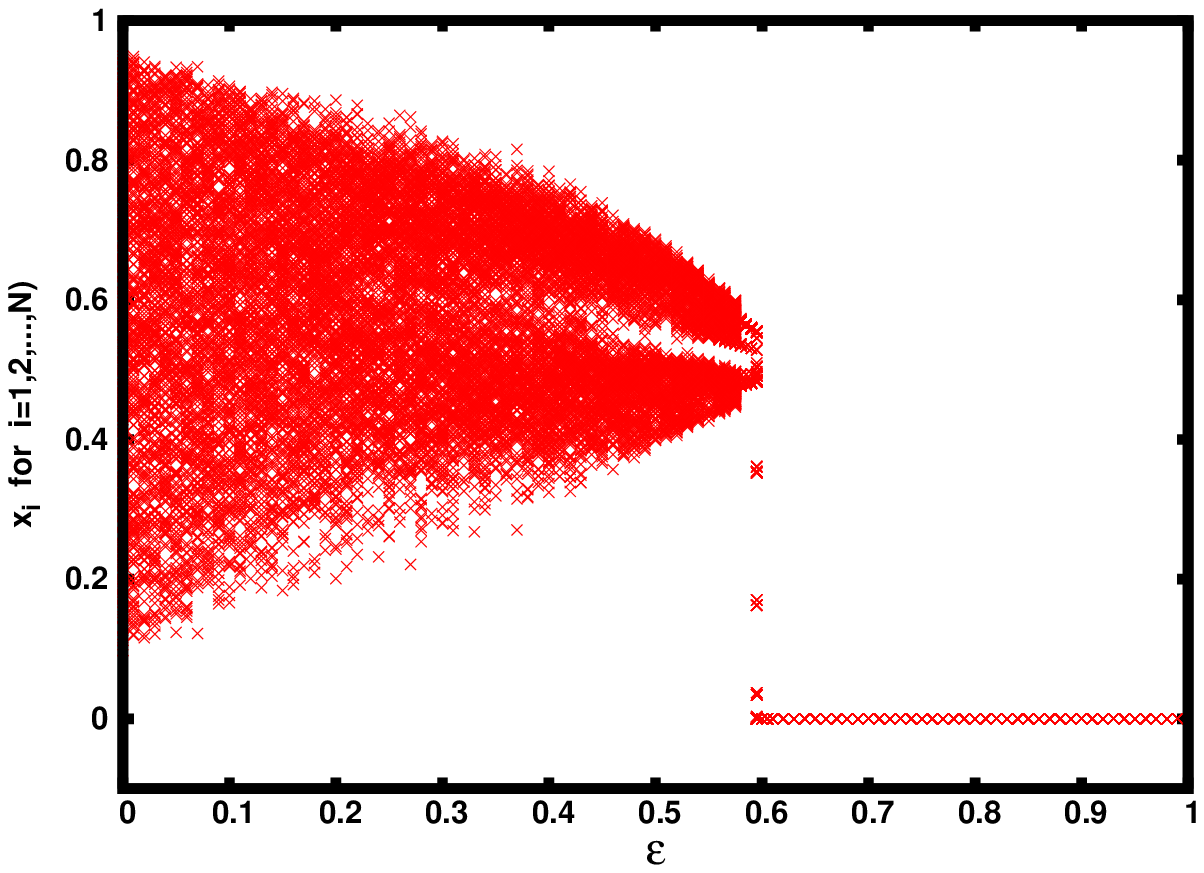}}
\caption{Bifurcation diagram showing values of $x_{n}^i$ with respect
  to coupling strength $\epsilon$, for coupled Tent maps with
  linear coupling ($q=1$ in Eqn.~1-2) and quadratic coupling ($q=2$ in
  Eqn.~1-2). Here the size of the lattice is $N=100$. In the figure we
  plot $x_n^i(i=1,...,100)$ over $n=1,...,10$ iterations (after a
  transience time of $5000$) for ten different initial conditions.}
\label{fig: bif13}
\end{figure}

\begin{figure}[]
\centering
\subfloat[Linear coupling]{\includegraphics[width=0.5\textwidth, height=50mm]{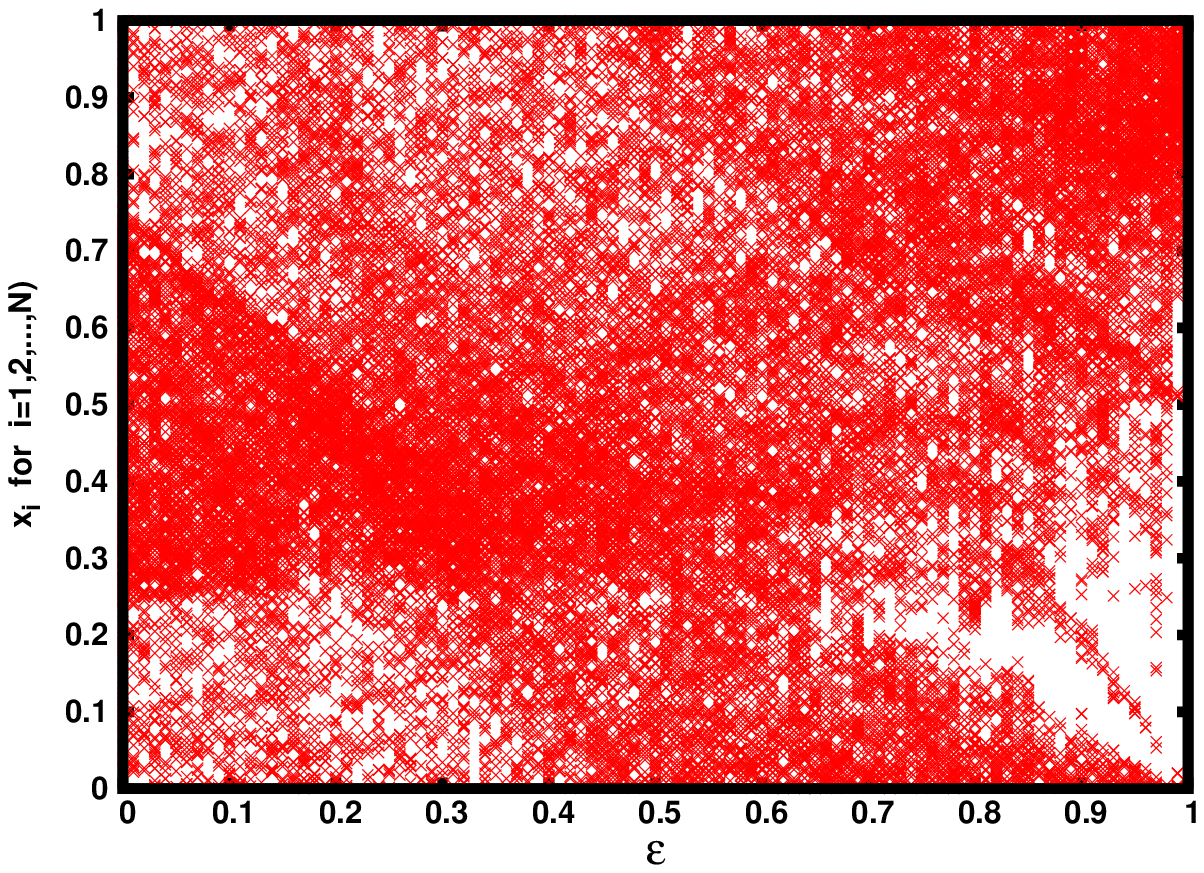}}
\subfloat[Quadratic Coupling]{\includegraphics[width=0.5\textwidth, height=50mm]{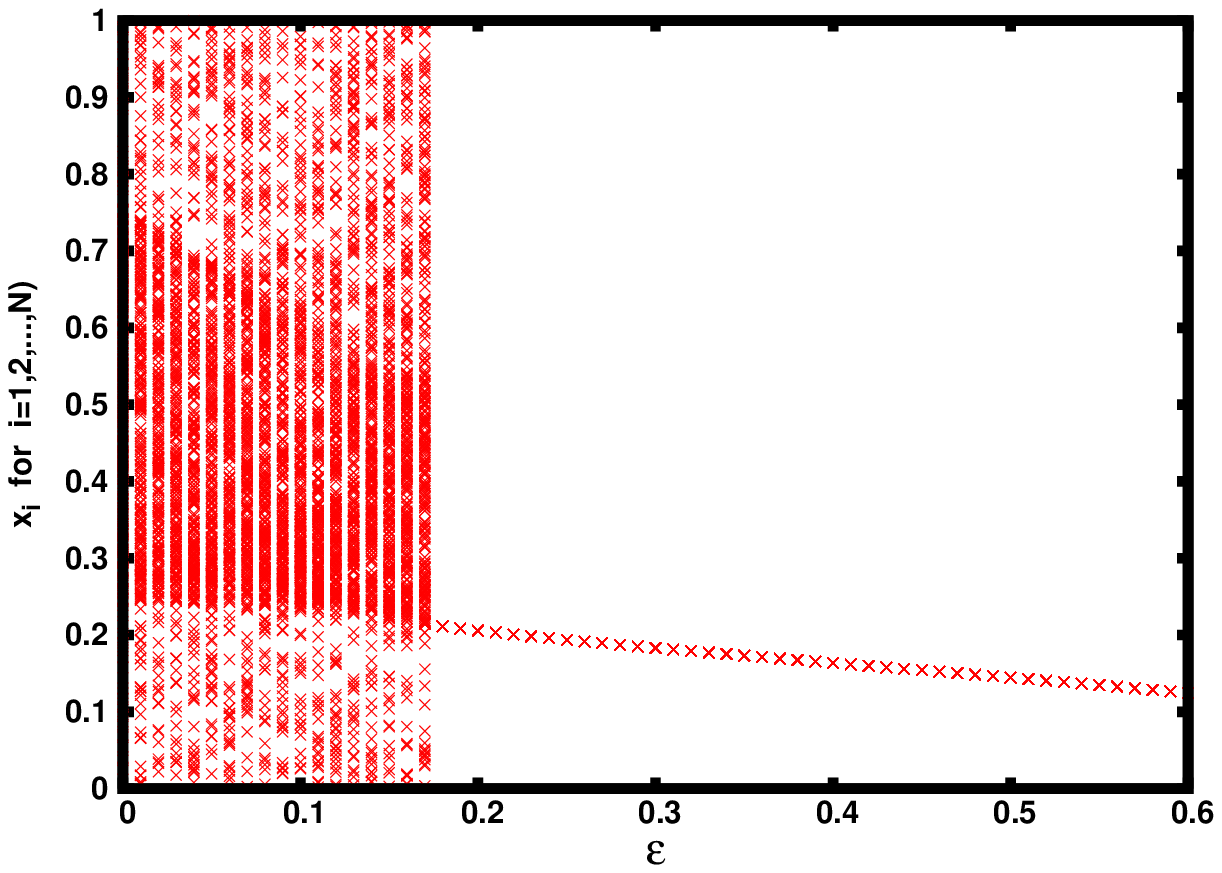}}
\caption{Bifurcation diagram showing values of $x_{n}^i$ with respect
  to coupling strength $\epsilon$, for coupled Sine circle maps with
  linear coupling ($q=1$ in Eqn.~1-2) and quadratic coupling ($q=2$ in
  Eqn.~1-2). Here the size of the lattice is $N=100$. In the figure we
  plot $x_n^i(i=1,...,100)$ over $n=1,...,10$ iterations (after a
  transience time of $5000$) for ten different initial conditions.}
\label{fig: bif14}
\end{figure}

We choose the chaotic Tent Map with parameter value $r=1.9$ as the
local dynamics. Results displayed in Figure \ref{fig: bif13}
clearly depict that while linear-coupled lattices of chaotic Tent maps
do not yield any stable spatiotemporal fixed point regimes, the
spatio-temporal fixed point state gains stability under nonlinear
coupling.

Lastly, we take  {\em Sine circle map} as local dynamics which has following equation:
$$ x_{n+1} = x_n + \omega - \frac{K}{2\pi}sin(2\pi x_n) \quad \quad (mod)1$$

For parameter values $\omega = 0.5$ and $K=3$ it is a chaotic function. Results presented in figure-\ref{fig: bif14} clearly show the
enhanced regularities in dynamics when we take nonlinear coupling.

\section*{Conclusions}

  In conclusion we have investigated the spatiotemporal behaviour of lattices
  of coupled chaotic logistic maps, where the coupling between sites
  has a nonlinear form. We showed that the stable range of the
  spatiotemporal fixed point state is significantly enhanced for
  increasingly nonlinear coupling. We demonstrated this through
  numerical simulations and linear stability analysis of the
  synchronized fixed point. Also, we showed that these results also
  hold in coupled map lattices where the nodal dynamics is given by
  the Gauss Map, Sine Circle Map and the Tent Map.

  The results here can be put in a following broader perspective: it
  is of considerable interest to find out what form of coupling and
  connection topologies allows the spatiotemporal fixed point of a
  collection of strongly chaotic elements to gain stability. This is
  of interest in the context of control, in both human engineered
  systems and in extended complex systems arising in the natural world
  \cite{ref03,ref06}. For instance, earlier studies had shown that
  dynamical random rewiring of links, namely changing the form of the
  connectivity matrix, stabilized the spatiotemporal fixed point in
  networks of chaotic elements \cite{ref02}. Here we show an alternate
  route to achieving spatiotemporal regularity, by demonstrating how
  in a regular lattice of chaotic maps we can achieve a stable
  synchronized fixed point by nonlinear coupling forms.

\subsection*{Acknowledgement}
I am deeply thankful to my masters project guide Prof. Sudeshna Sinha
for her guidance, help and encouragement. Also I thank to Mr. Anshul
Choudhary and Anshu Gupta for their fruitful discussions.
I dedicate this work to my family. 
\bibliographystyle{unsrt}


\bibliography{logistic}

\end{document}